\documentclass[12pt]{iopart}
\usepackage[usenames,dvipsnames]{pstricks}
\usepackage[T1]{fontenc}
\usepackage[latin9]{inputenc}
\usepackage{lmodern,times}
\usepackage{graphicx}
\usepackage{fancybox}
\usepackage{units}
\usepackage{harvard}
\usepackage{bm,iopams,amssymb,amsfonts}
\newcommand{\REM}[1]{}
\eqnobysec
\begin{document}
\bibliographystyle{jphysicsB}

\title{Finite--size effects in the spherical model of finite thickness}
%\title{Mean spherical model with Dirichlet boundary conditions}
\author{H. Chamati
\footnote{On leave from Institute
of Solid State Physics, Bulgarian Academy of Sciences,
72 Tzarigradsko Chauss\'ee, 1784 Sofia,
Bulgaria}}
\address{Universit\"at Duisburg--Essen,
Fachbereich Physik,
%Lotharstra\ss e 1,
47048 Duisburg}
\ead{hassan.chamati@uni-due.de}

\begin{abstract}
A detailed analysis of the finite--size effects on the bulk critical
behaviour of the $d$--dimensional mean spherical model confined to a
film geometry with finite thickness $L$ is reported. Along the finite
direction different kinds of boundary conditions are applied: periodic $(p)$,
antiperiodic $(a)$ and free surfaces with Dirichlet $(D)$, Neumann
$(N)$ and
a combination of Neumann and Dirichlet $(ND)$ on both surfaces. A
systematic method for the evaluation of the finite--size corrections
to the free energy for the different types of boundary conditions is
proposed. The free energy density and the equation for the spherical
field are computed for arbitrary $d$. It is found, for $2<d<4$,
that the singular
part of the free energy has the required finite--size scaling form at the
bulk critical temperature only for $(p)$ and $(a)$. For the remaining
boundary conditions the standard finite--size scaling hypothesis
is not valid.
% at a
%shifted, due to the surfaces, critical temperature.
At $d=3$, the
critical amplitude of the singular part of the free energy (related to the so called Casimir amplitude) is
estimated. We obtain $\Delta^{(p)}=-2\zeta(3)/(5\pi)=-0.153051\ldots$,
$\Delta^{(a)}=0.274543\ldots$ and $\Delta^{(ND)}=0.01922\ldots$,
%thus
implying a fluctuation--induced attraction between the surfaces for
$(p)$ and repulsion in the other two cases. For $(D)$ and $(N)$ we find a
logarithmic dependence on $L$.
\end{abstract}
%\pacs{64.60.--i,64.60.F--,75.40.--s}
\submitto{J. Phys. A: Math. Theor.}
\maketitle

\section{Introduction}
Over the last 30 years there has been an increasing interest in the
critical behaviour of systems (fluids, magnets \ldots) confined between two infinite parallel
plates i.e. films. Such systems can be regarded as $d$--dimensional
generalization
of two $(d-1)$--dimensional walls separated by matter of thickness $L$.
The so called thermodynamic Casimir effect
related to fluctuation induced long--ranged forces between the plates
was, and still is, the central theme of these investigations
both experimentally and theoretically\footnote{The literature on
the thermodynamic Casimir effect can be found in
\cite{grueneberg2008} and references therein.}. This effect was first
predicted by Fisher and de Gennes \cite{fisher1978} in their
investigation on a confined critical binary liquid at its bulk
demixing point. They showed that
the reduced free energy per unit area contains a term of the
form $\Delta^{(\tau)}L^{-d+1}$, where
%$L$ is the thickness of the film and
$\Delta^{(\tau)}$ is the so called Casimir amplitude by analogy with the
Casimir effect in vacuum fluctuations of the electromagnetic field between
two metallic plates \cite{casimir1948}. It is a
universal quantity that depends on the bulk universality class and the
boundary conditions $(\tau)$ imposed on the confining walls
\cite{privman1984,singh1985a}.
The different types of boundary conditions are in turn related to
distinct universality classes of surface critical behaviour depending
on the behaviour of the order parameter at the surfaces bounding the
system and some additional surface properties \cite{diehl1997}.

In general, according to \citeasnoun{privman1984} and \citeasnoun{singh1985a} the singular part of
the free--energy density of a finite (in one or more directions)
$d$--dimensional system with linear size $L$,
near the bulk critical point $T_c$, may be expressed in the form
\begin{equation}\label{PF}
f^{(\tau)}_{s,d}(t,h;L)\approx
L^{-d}Y^{(\tau)}(c_1tL^{1/\nu},c_2hL^{\Delta/\nu}),
\end{equation}
where $t$ and $h$ are related to the temperature, $T$, and the external
magnetic field, $H$, via
\begin{equation}
t=\frac{T-T_c}{T_c}, \qquad h=\frac{H}{k_BT}.
\end{equation}
The arguments of $Y^{(\tau)}(x_1,x_2)$ are appropriate scaled variables, $\nu$
and $\Delta$ are the usual critical exponents, while all the details
of the system are incorporated in the non--universal quantities $c_1$
and $c_2$. Then the function $Y^{(\tau)}(x_1,x_2)$ is a universal scaling
function, whose exact expression depends upon the number of
finite directions, the bulk universality class and the boundary
conditions $(\tau)$ to which the system is subjected. For a system
confined to a film geometry, at $T=T_c$, the Casimir amplitude coincides
with the critical amplitude of the singular part of the free energy
density, i.e.
\begin{equation}
\Delta^{(\tau)}=Y^{(\tau)}(0,0).
\end{equation}

The spherical model of Berlin and Kac \cite{berlin1952} was initially designed to
mimic the critical properties of the Ising model. It has been obtained
by requiring the spins to be continuous variables subject to a global constraint
(the sum of the squares of spins at each lattice site is equal to the total
number of sites $N$) rather than a local one (the square of the spin at
each site is exactly 1). Later it was shown
that the free energy of this model can be obtained as a limiting case
of that of the Heisenberg model with infinite number of spin components
\cite{stanley1968}. The equivalence between the Heisenberg model with
infinite spin components and the spherical model remains valid for
finite systems, as long as one considers boundary conditions that
preserve the translational invariance of the lattice
\cite{knops1973}. \citeasnoun{lewis1952} proposed to
simplify the spherical model by requiring the global constraint of
Berlin and Kac to be obeyed in the sense of an ensemble average. This
model is generally known as the \textit{mean spherical model}.

Because of its exact solubility
%and direct relevance to real physical situations
the ferromagnetic mean spherical model has been extensively
used to gain insights into
the critical properties of finite (in one or more directions)
systems\footnote{For an extensive list of literature see
\cite{barber1973,singh1985a,privman1990,chamati1998,brankov2000}
and references therein.}.
Periodic boundary conditions, implying ferromagnetic interactions of
spins belonging to both boundaries on the finite directions,
%the order parameter at both ends of the finite directions,
have been by far the
most used boundary conditions. These allow for analytic
treatment of problems related to the finite--size scaling theory
\cite{privman1991,brankov2000}.
In addition to periodic boundary
conditions, antiperiodic boundary conditions
%(antiferromagnetic interaction involving only spins on opposite
%boundaries along the finite directions)
have also
been used \cite{barber1973,singh1985b} to investigate the
finite--size scaling properties of the ferromagnetic mean spherical
model. The aforementioned boundary
conditions do not break the translational invariance of the model in
the absence of a magnetic field. Much less has been achieved in the
case of nonperiodic (free) boundary conditions. These are believed to be more relevant
to real systems, especially to systems confined between parallel
plates. The ferromagnetic mean spherical model of finite thickness has been
investigated in the case of Dirichlet boundary conditions
%(vanishing order parameter at the plates bounding the system)
by
\citeasnoun{barber1973}, \citeasnoun{barber1974a},
\citeasnoun{danchev1997}, \citeasnoun{chen2003} and
\citeasnoun{dantchev2003}, and Neumann boundary
conditions
%(surface spins interacting with the plates bounding the system)
by \citeasnoun{barber1974b}, \citeasnoun{danchev1997} and \citeasnoun{dantchev2003}.
%The authors of
%\cite{barber1973,barber1974a} using a method originally
%devised in \cite{barber1973} investigated the scaling properties of
%the mean spherical model with finite thickness.
\citeasnoun{barber1973} and \citeasnoun{barber1974a} investigated the scaling
properties of the mean spherical model with a finite thickness using a
method originally devised by \citeasnoun{barber1973}.
They obtained explicit
forms of the equation for the spherical field at three, four
and five dimensions, separately. The method was extended to the
study of the finite--size effects in the case of Neumann and
Neumann--Dirichlet boundary conditions by \citeasnoun{danchev1997}.
\citeasnoun{chen2003} argued
that the results of \citeasnoun{barber1973} and \citeasnoun{barber1974a} for Dirichlet boundary
conditions were incorrect at three dimensions far from the critical
point. Later, the equation for the spherical field of \citeasnoun{barber1973} was
rederived by \citeasnoun{dantchev2003}. It should be mentioned, however that the derivation of
\citeasnoun{dantchev2003} is based on an improved method of
\citeasnoun{barber1973}. On the other hand, in
\cite{barber1973,barber1974a,danchev1997,dantchev2003} the free energy
density for the different kinds of boundary conditions was obtained by
integrating the equation for the spherical field leading to
complicated integral representations.

In the present paper we propose a different method to treat the finite
size effects in the mean spherical model of finite thickness. The
method is a generalization of that devised by
\citeasnoun{singh1985a} for periodic boundary conditions. It applies to
antiperiodic, Dirichlet, Neumann boundary conditions imposed on both
surfaces bounding the system and a combination of Dirichlet and
Neumann boundary conditions on each surface.
The method is quite general and is used for arbitrary dimension.
Furthermore, the finite--size contributions to the free energy
density are obtained directly form the corresponding
general expression rather than through integration of the equation for
the spherical field. We anticipate here that from our results
we recover the particular cases of
\citeasnoun{barber1973}, \citeasnoun{barber1974a},
\citeasnoun{danchev1997} and \citeasnoun{dantchev2003} but not the results of
\citeasnoun{chen2003}. We will return to these points later in the paper.

The rest of the paper is structured as follows: In Section
\ref{smodel} we define the model and present the expressions for the
free energy density and the equation for the spherical field for the
different kinds of boundary conditions. In Section \ref{periodic} we
present in detail how the method applies to the case of periodic boundary
conditions and compare our results with those available in the
literature. In Sections \ref{antiperiodic} through \ref{smixed} we
extend the method for arbitrary dimensions to the other boundary
conditions and compare with the results obtained by other authors
using different methods. Finally in Section \ref{discussion}
we summarize our results.

\section{The model}\label{smodel}
We consider the mean spherical model on a $d$--dimensional lattice confined to a
film geometry i.e. infinite in $d-1$ dimensions and of finite
thickness $L$ in the remaining dimension with volume
$V=L\times\infty^{d-1}$. The linear size of the lattice is measured in
units of the lattice constant, which will be taken equal to 1. The model is defined through
\begin{equation}\label{model}
\mathcal{H}=-\case{1}{2}\sum_{ij}J_{ij}s_is_j-H\sum_is_i
+\case{1}{2}\mu\sum_{i}s_i^{\ 2}
\end{equation}
where $s_i\equiv s_i({\bm r},z)$, the spin at site $i$,
is a continuous real variable ($-\infty\leq s_i\leq\infty$) with
coordinates ${\bm r}$ in the $d-1$ infinite ``parallel'' planes
and $z$ in the finite lateral direction. $J_{ij}$ is the interaction
matrix between spins at sites $i$ and $j$, and $H$ is an ordering
external magnetic field. Finally the field $\mu$ is
introduced so as to ensure the spherical constraint
\begin{equation}
\sum_i\langle s_i^2\rangle=N,
\end{equation}
where $\langle\cdots\rangle$
denotes standard thermodynamic averages computed with
the Hamiltonian $\mathcal{H}$ and $N$ the total number of spins on the
lattice.

Along the finite $z$ direction we impose different kinds of boundary
conditions which we will denote collectively by $\tau$.
For a lattice system this means:

\begin{itemize}
\item[$(p)$] periodic: $s({\bm r},1)=s({\bm r},L+1)$;

\item[$(a)$]  antiperiperiodic: $s({\bm r},1)=-s({\bm r},L+1)$;

\item[$(D)$]  Dirichlet: $s({\bm r},0)=s({\bm r},L+1)=0$;

\item[$(N)$]  Neumann: $s({\bm r},0)=s({\bm r},1)$ and $s({\bm r},L)=s({\bm r},L+1)$;

\item[$(ND)$]  Neumann--Dirichlet: $s({\bm r},0)=s({\bm r},1)$ and $s({\bm r},L+1)=0$.
\end{itemize}

For nearest
neighbour interactions and under the above boundary conditions the
Hamiltonian (\ref{model}) may be diagonalized by plane waves parallel
to the confining plates and
appropriate eigenfunctions $\varphi_n^{(\tau)}(z)$ using the representation
\begin{equation}
s_i({\bm r},z)=\sum_{n}
\int_0^{2\pi}\frac{d q_1}{2\pi}
\cdots\int_0^{2\pi}\frac{d q_{d-1}}{2\pi}
s_{\bm q,n}
e^{i{\bm q}\cdot{\bm r}}\varphi_n^{(\tau)}(z),
\end{equation}
where the integrals over $q_i$s, the components of $\bm{q}$,
are restricted to the first Brillouin
zone of the hypercubic lattice of dimension $d-1$.
The orthogonal eigenfunctions read
\numparts
\begin{equation}
\fl\varphi_n^{(p)}(z)=\frac1{\sqrt{L}}\exp\left[i \frac{2\pi}L nz\right],
\qquad n=0,\cdots,L-1;
\end{equation}
\begin{equation}
\fl\varphi_n^{(a)}(z)=\frac1{\sqrt{L}}\exp\left[i\frac{2\pi}L\left(n+\case{1}{2}\right)z\right],
\qquad n=0,\cdots,L-1;
\end{equation}
\begin{equation}
\fl\varphi_n^{(D)}(z)=\sqrt{\frac2{L+1}}\sin\left[\frac{\pi}{L+1}(n+1)z\right],
\qquad n=0,\cdots,L-1;
\end{equation}
\begin{equation}
\fl\varphi_n^{(N)}(z)=\left\{
\begin{array}{ll}
\displaystyle L^{-1/2} ,               &n=0,   \\[.4cm]
\sqrt{\frac2L}\cos\left[\frac{\pi}Lnz\right] ,
\qquad &n=1,\ldots,L-1
\end{array}
\right.;
\end{equation}
\begin{equation}
\fl\varphi_n^{(ND)}(z)=\frac2{\sqrt{2L+1}}
\cos\left[\frac{2\pi}{2L+1}\left(n+\case{1}{2}\right)z\right];
\qquad n=0,\cdots,L-1.
\end{equation}
\endnumparts
The eigenmodes associated with the above eigenfunctions are given by
\numparts
\begin{equation}
\omega_n^{(p)}=-2+2\cos\left[\frac{2\pi}Ln\right];
\end{equation}
\begin{equation}
\omega_n^{(a)}=-2+2\cos\left[\frac{2\pi}L\left(n+\case{1}{2}\right)\right];
\end{equation}
\begin{equation}
\omega_n^{(D)}=-2+2\cos\left[\frac{\pi}{L+1}(n+1)\right];
\end{equation}
\begin{equation}
\omega_n^{(N)}=-2+2\cos\left[\frac{\pi}Ln\right];
\end{equation}
\begin{equation}
\omega_n^{(DN)}=-2+2\cos\left[\frac{2\pi}{2L+1}\left(n+\case{1}{2}\right)\right].
\end{equation}
\endnumparts
Those corresponding to the diagonalized interaction matrix are
$\lambda_n^{(\tau)}=2+\omega_0^{(\tau)}$.

At zero field, the free energy density of the mean spherical model under the above
boundary conditions imposed along the finite lateral size has the general
expression
(obtained via the Legendre transformation)
\begin{equation}\label{freee}
\fl\beta F_d^{(\tau)}(T,L;\phi)=\frac12\ln K
+\frac1{2L}\sum_{n=0}^{L-1}U_{d-1}(\phi+\omega_0^{(\tau)}
-\omega_n^{(\tau)})
-\frac12K(\phi+\omega_0^{(\tau)})-K,
\end{equation}
where $K=\beta J=J/k_BT$
and
\begin{equation}
U_d(z)\!=\!\int_0^{2\pi}\frac{d\theta_1}{2\pi}
\cdots\int_0^{2\pi}\frac{d\theta_d}{2\pi}
\ln\left[z+2\sum_{i=1}^d(1-\cos\theta_i)\right].
\end{equation}
In (\ref{freee}), the shifted spherical field, defined by
$$
\phi=\mu/J-\lambda_0^{(\tau)},
$$
is the solution of the spherical constraint
\begin{equation}\label{phi}
K=\frac1L\sum^{L-1}_{n=0}W_{d-1}(\phi+\omega_0^{(\tau)}-\omega_n^{(\tau)}).
\end{equation}
with the bulk function
\begin{equation}\label{calw}
W_d(z)=\int_0^{2\pi}\frac{d\theta_1}{2\pi}
\cdots\int_0^{2\pi}\frac{d\theta_d}{2\pi}
\frac1{z+2\sum_{i=1}^d(1-\cos\theta_i)},
\end{equation}
whose asymptotic behaviour has been studied in
considerable detail for $z\in\mathbb{C}$ in \cite{barber1973}.

The bulk limit is obtained from (\ref{phi}) by letting the lateral size $L$ go to
infinity.
This allows us to investigate the thermodynamics of the ferromagnetic
mean spherical model in the thermodynamic limit. Here we shall not enter
into the investigation of the thermodynamic properties (the interested reader
may refer to \cite{pathria1996}). It is worth mentioning that the present
model undergoes a continuous phase transition at a bulk critical point determined by
\begin{equation}
K_{c,d}=W_d(0),
\end{equation}
for $d>2$.
On the
other hand for $d>4$ the model exhibits mean--field--like critical
behaviour.
%In the following we will consider only dimensions in the interval $2<d<4$.
Recently an efficient method to estimate the Watson
integral $W_d(0)$ and the associated logarithmic integral for
$d$--dimensional hypercubic lattice has been proposed by \citeasnoun{joyce2001}.

Before embarking into the investigation of the finite--size effects
for the different boundary conditions a few comments are in order:

(1) For the evaluation of the finite--size contributions to the bulk
expressions of the free energy and the equation for the spherical
field the case of periodic boundary conditions is the simplest. A
powerful method to treat these sums was proposed in
\cite{singh1985a}. It was found that in this case the correlation
length $\xi_L$ of the finite system is given by
$\xi_L=\phi^{-1/2}$ \cite{singh1987}.

(2) The same method was successfully extended to the case of antiperiodic boundary
conditions
\cite{singh1985b}. However, the calculation were done in the absence
of the lowest mode $\omega_0^{(\tau)}$ in (\ref{freee}) and (\ref{phi}).
Later by investigating the correlation function
\cite{allen1993}, it was found that
the correlation length $\xi_L$ and the root square of spherical field $\phi$ are no more
connected by a simple relation as it is the case for periodic boundary
conditions. It was suggested that the correlation
length is in fact related to the solution of the equation for
spherical field shifted by the asymptotic behaviour of
$\omega_0^{(\tau)}$ i.e. $\frac{\pi^2}{L^2}$.

(3) The remaining boundary conditions have attracted less attention
\cite{barber1973,barber1974a,barber1974b,danchev1997,brankov2000,chen2003,dantchev2003}.
Apart from the paper by \cite{chen2003}, all investigations were
specialized to $d=3$ using an approach based on \cite{barber1973}. For arbitrary $d$
and Dirichlet boundary conditions, a recent work by
\citeasnoun{chen2003} proposed a different method and claimed that the
results of \cite{barber1973}, obtained at $d=3$, were incorrect.

Here we generalize the method of \cite{singh1985a} for periodic
boundary conditions to the other boundary conditions. We investigate
the finite--size effects of the free energy density (\ref{freee}) and the equation
for the spherical field (\ref{phi}) in each case for arbitrary
dimension
and comment on the results of the aforementioned papers.

\section{Periodic boundary conditions}\label{periodic}
For the sake of completeness we will derive here the relevant expressions
for the mean spherical model of finite thickness under periodic
boundary conditions for arbitrary dimensionality. The derivation is adapted from
\cite{singh1985a} (see also \cite{chamati1998}).
The explicit expressions for the free energy density (\ref{freee})
and the equation for the spherical field (\ref{phi}) are given by
\numparts
\begin{equation}\label{fpbc}
\fl\beta F_d^{(p)}(T,L)=\frac12\ln K
+\frac1{2L}\sum_{n=0}^{L-1}
U_{d-1}\left(\phi+2\left[1-\cos\left(\frac{2\pi}Ln\right)\right]\right)
-\frac12K\phi-K,
\end{equation}
and
\begin{equation}\label{phipbc}
K=\frac1L\sum^{L-1}_{n=0}
W_{d-1}\left(\phi+2\left[1-\cos\left(\frac{2\pi}Ln\right)\right]\right),
\end{equation}
respectively. In this case $\omega_0^{(p)}=0$.
\endnumparts

Using the integral representations
\numparts
\begin{equation}\label{intrep1}
z^{-1}=\int_0^\infty e^{-zt}
\end{equation}
and
\begin{equation}\label{intrep2}
\ln z = \int_0^\infty \frac{dt}t\left[e^{-t}-e^{-zt}\right].
\end{equation}
\endnumparts
we can separate the expressions for the free energy density (\ref{fpbc})
and the equation for the spherical field (\ref{phipbc}) into the
corresponding bulk expressions and the associated finite--size contributions.
Let us illustrate how this works for the equation for the spherical
field. With the aid of (\ref{intrep1}) the sum entering this equation can be written as:
\begin{eqnarray}\label{spbc}
\fl\mathcal{S}_{d,L}^{(p)}(\phi)&=\sum^{L-1}_{n=0}
W_{d-1}\left[\phi+2\left(1-\cos\frac{2\pi}{L}n\right)\right]\nonumber\\
\fl&=\int_0^{2\pi}\frac{d\theta_1}{2\pi}
\cdots\int_0^{2\pi}\frac{d\theta_{d-1}}{2\pi}
\int_0^\infty dz\exp\left\{-z\left[\phi +2
+2\sum_{i=1}^{d-1}(1-\cos\theta_i)\right]\right\}Q_L^{(p)}(2z)\nonumber\\
\fl&
\end{eqnarray}
with
\begin{equation}\label{qofz}
Q_L^{(p)}(z)=\sum_{n=0}^{L-1}\exp\left[z\cos\frac{2\pi}L n\right].
\end{equation}
Notice that the summand in (\ref{qofz}) is a periodic function of
period $2\pi$.
Further we use (a generalization of) the Poisson summation formula, namely
\begin{equation}
\sum_a^bf(n)=\sum_{l=-\infty}^\infty\int_a^b e^{2\pi il n}f(n)dn
+\frac12f(a)+\frac12f(b).
\end{equation}
to get the identity
\begin{equation}\label{identityp}
Q_L^{(p)}(z)=\sum_{n=0}^{L-1}\exp\left[z\cos\frac{2\pi}Ln\right]
=L\sum_{l=-\infty}^\infty I_{Ll}(z),
\end{equation}
where $I_\nu(x)$ stands for the modified Bessel function of the first
kind \cite{abramowitz1972}.

Substituting (\ref{identityp}) into (\ref{spbc}) we obtain
\begin{eqnarray}
\fl\mathcal{S}_{d,L}^{(p)}(\phi)&=L\sum_{l=-\infty}^\infty\int_0^\infty dz
e^{-z\phi}
\left[e^{-2z}I_0(2z)\right]^{d-1}e^{-2z}I_{Ll}(2z)\nonumber\\
\fl&=LW_d(\phi)
+2L\sum_{l=1}^\infty\int_0^\infty dz e^{-z\phi}
\left[e^{-2z}I_0(2z)\right]^{d-1}e^{-2z}I_{Ll}(2z),
\end{eqnarray}
where we have used the integral representation
\begin{equation}
W_d(\phi)=
\int_0^\infty dz e^{-z\phi}\left[e^{-2z}I_0(2z)\right]^d
\end{equation}
and the relation $I_{-2p}(x)=I_{2p}(x)$ \cite{abramowitz1972}.

For large $L$, using the asymptotic expansion \cite{singh1985a}
\begin{equation}
I_\nu(x)=\frac{e^{x-\nu^2/2x}}{\sqrt{2\pi
x}}\left[1+\frac1{8x}+\frac{9-32\nu^2}{2!(8x)^2}
+\cdots\right],
\end{equation}
after some straightforward steps, keeping only leading terms in
$L^{-1}$, we get
\begin{equation}\label{spres}
\mathcal{S}^{(p)}_{d,L}(\phi)=LW_d\left(\phi\right)
+L\frac4{(4\pi)^{\frac d2}}\phi^{\frac d2-1}
\sum_{l=1}^\infty
\frac{K_{\frac d2-1}\left(lL\sqrt{\phi}\right)}
{(\frac12lL\sqrt{\phi})^{\frac d2-1}},
\end{equation}
where $K_{\nu}(x)$ is the modified Bessel function of the second kind
\cite{abramowitz1972}.

Collecting the above results the equation for the spherical field reads
\begin{equation}\label{phip}
K=W_d(\phi)
+\frac4{(4\pi)^{\frac d2}}\phi^{\frac d2-1}
\sum_{l=1}^\infty
\frac{K_{\frac d2-1}\left(lL\sqrt{\phi}\right)}
{(\frac12lL\sqrt{\phi})^{\frac d2-1}}.
\end{equation}
The finite--size behaviour of this equation has been studied in great detail
by \citeasnoun{chamati1998} for arbitrary $d$ for different geometries
including questions like dimensional crossover, finite--size shift of
the bulk critical temperature, etc. In particular it was found that
the finite--size shift obeys the predictions of the finite--size scaling.
For $2<d<4$, using the asymptotic behaviour
\begin{equation}\label{aswb}
W_d(z)= W_d(0)
+\frac1{(4\pi)^{d/2}}\Gamma\left[\frac{2-d}2\right]z^{(d-2)/2}+O\left(z^{(d-1)/2}\right),
\end{equation}
equation (\ref{phip}) takes the scaling form
\begin{equation}\label{kappap}
\varkappa=
\frac{y^{d-2}}{(4\pi)^{d/2}}
\left[\left|\Gamma\left[\frac{2-d}2\right]\right|
-4
\sum_{l=1}^\infty
\frac{K_{\frac d2-1}\left(ly\right)}{(\frac12ly)^{\frac d2-1}}\right],
\end{equation}
where we have introduced the scaling variable $y=L\sqrt{\phi}$ and
$\varkappa=L^{1/\nu}(K_{c,d}-K)$ with $\nu=(d-2)^{-1}$ -- the critical
exponent measuring the divergence of the correlation length.
Consequently we have a solution of the form
$\xi_L=\phi^{-1/2}=Lf_{{p}}(\varkappa)$, where $f_{(p)}$ is a
universal scaling function. For arbitrary $d$ the nature of the
scaling function $f_{(p)}(\varkappa)$ can be determined only
numerically (see e.g. \cite{chamati1998}).
For the particular case $d=3$, equation (\ref{kappap}) takes a simple form
\begin{equation}
2\pi\varkappa=\ln2\sinh\frac y2.
\end{equation}
The solution of this equation leads to the universal scaling function
\begin{equation}\label{sfp}
y=g_{(p)}(\varkappa)=2\ \mathrm{arcsinh\left(\case12e^{2\pi\varkappa}\right)}
\end{equation}
At the critical point, $\varkappa=0$, we obtain the critical amplitude
of the finite--size correlation length $\xi_L$:
\begin{equation}\label{pa}
y_0=g_{(p)}(0)=2\ln\frac{1+\sqrt{5}}2.
\end{equation}

The finite--size contributions to the free energy for any $d$ can be accounted for
by using the integral representation (\ref{intrep2}).
The aim here is to transform the sum
\begin{eqnarray}\label{pper}
\fl\mathcal{P}_{d,L}^{(p)}(\phi)&
=\sum_{n=0}^{L-1}
U_{d-1}\left[\phi+2\left(1-\cos\frac{2\pi}Ln\right)\right]\nonumber\\
\fl&
=\sum_{n=0}^{L-1}\int_0^{2\pi}\frac{d\theta_1}{2\pi}
\cdots\int_0^{2\pi}\frac{d\theta_{d-1}}{2\pi}
\ln\left[\phi+2\left(1-\cos\frac{2\pi}Ln\right)
+2\sum_{i=1}^{d-1}(1-\cos\theta_i)\right]
\end{eqnarray}
into a more tractable form suitable for analytic treatment.
After some straightforward algebra
along the lines explained above,
including the use of the identity (\ref{identityp}), we arrive at
\begin{equation}\label{pp}
\mathcal{P}_{d,L}^{(p)}(\phi)=LU_d(\phi)
-L\frac4{(4\pi)^{\frac d2}}\phi^{\frac d2}
\sum_{l=1}^\infty\frac{ K_{\frac d2}
\left(lL\sqrt{\phi}\right)}{\left(\frac12lL\sqrt{\phi}\right)^{\frac d2}},
\end{equation}

Using (\ref{pp}) for the large $L$ asymptotic behaviour of the
free energy density (\ref{fpbc}) we get
\begin{equation}\label{fpf}
\beta F_d^{(p)}(T,L;\phi)=\beta F_d(T;\phi)
-\frac2{(4\pi)^{\frac d2}}\phi^{\frac d2}
\sum_{l=1}^\infty\frac{ K_{\frac d2}
\left(lL\sqrt{\phi}\right)}{\left(\frac12lL\sqrt{\phi}\right)^{\frac d2}},
\end{equation}
where
%$F_{d}(T;\phi)=\lim_{L\to\infty}F_d^{(\tau)}(T,L;\phi)$
\begin{equation}
F_{d}(T;\phi)=\lim_{L\to\infty}F_d^{(\tau)}(T,L;\phi)
             =\frac1{2\beta}\left[\ln K+U_d(\phi)-K\phi-2K\right]
\end{equation}
is the bulk free energy density.

For $2<d<4$, in the vicinity of the bulk critical point, we have the
expansion
\begin{equation}\label{uss}
U_d(z)= U_d(0)+K_{c,d}z
+\frac1{(4\pi)^{d/2}}\frac2d\Gamma\left[\frac{2-d}2\right]z^{d/2}
+O(z^{(d+1)/2}).
\end{equation}
Then the singular part of the free energy density (\ref{fpf}) takes the scaling form
\begin{equation}\label{singp}
\fl\beta f_{s,d}^{(p)}(y,\varkappa)=
\frac12L^{-d}\left[\varkappa y^2+
\frac{2y^d}{(4\pi)^{d/2}}\left(\frac1d\Gamma\left[\frac{2-d}2\right]
-2
\sum_{l=1}^\infty\frac{ K_{\frac d2}
\left(ly\right)}{\left(\frac12 ly\right)^{\frac
d2}}\right)\right],
\end{equation}
where $y$ is the solution of (\ref{kappap}). Thus, the scaling behaviour of the free
energy density is
consistent with the finite--size scaling hypothesis (\ref{PF}).
The Casimir amplitude for $d=3$ i.e. the critical
amplitude of the singular part of the free energy density can be
computed analytically,
having in mind the solution (\ref{pa}), with the aid of
polylogarithmic identities \cite{sachdev1993}.
The result is \cite{sachdev1993,danchev1998}:
\begin{equation}\label{casper}
\Delta^{(p)}=-\frac{2\zeta(3)}{5\pi}=-0.153051\ldots.
\end{equation}
This value is compatible with estimations obtained for more realistic $O(n)$
models using different approaches. For more details we refer the
reader to \cite{grueneberg2008}.

As one would expect from previous studies on the mean spherical model,
the results obtained for periodic boundary conditions are in
conformity with the finite--size scaling hypothesis. In the following
we will extend the method described here to the other boundary
conditions. The generalization is somehow straightforward, however we
will see the appearance of some subtleties that need careful
consideration.

\section{Antiperiodic boundary conditions}\label{antiperiodic}
The explicit expressions for the free energy density (\ref{freee})
and the equation
for the spherical field (\ref{phi}) for arbitrary dimension read
\numparts
\begin{equation}\label{fabc}
\fl\beta F_d^{(a)}(T,L;\phi)=\frac12\ln K +\frac1{2L}\sum_{n=0}^{L-1}
U_{d-1}\left[\phi+\omega_0^{(a)}-\omega_n^{(a)}\right]
-\frac12K\left(\phi+\omega_0^{(a)}\right)-K,
\end{equation}
and
\begin{equation}\label{phiabc}
K=\frac1L\sum^{L-1}_{n=0}
W_{d-1}
\left[\phi+\omega_0^{(a)}-\omega_n^{(a)}\right].
\end{equation}
respectively. Here the lowest eigenmode $\omega_0^{(a)}\neq0$,
in contrast to the case of periodic boundary conditions.
\endnumparts

Unlike the analysis of \cite{singh1985b} we shall not omit
$\omega_0^{(\tau)}=-2+2\cos\frac{\pi}L$ from our equations, rather we will use the
combination $\sigma^{(a)}=\phi+\omega_0^{(a)}$ as a variable in our consideration of the
sums entering (\ref{fabc}) and (\ref{phiabc}). We will see
that this is crucial to our further treatment. Indeed, in our
notations, $\phi$ is expected to define the correlation length and no other
definition for this quantity is necessary as it has been suggested by
\citeasnoun{allen1993}, where the behaviour of order parameter correlation function
for the finite mean--spherical model with antiperiodic boundary conditions
has been investigated in detail.

The analysis of Section \ref{periodic} for periodic boundary
conditions can be applied to the sums
\numparts
\begin{eqnarray}\label{sapbc}
\mathcal{S}_{d,L}^{(a)}[\sigma^{(a)}]&=
\sum_{n=0}^{L-1}W_{d-1}
\left[\sigma^{(a)}-\omega_n^{(a)}\right]\\
\mathcal{P}_{d,L}^{(a)}[\sigma^{(a)}]&=
\sum_{n=0}^{L-1}U_{d-1}
\left[\sigma^{(a)}-\omega_n^{(a)}\right]\label{papbc}
\end{eqnarray}
\endnumparts
appearing in the left hand side of (\ref{fabc}) and (\ref{phiabc}),
respectively.
Now instead of the identity (\ref{identityp}) we use \cite{singh1985b}
\begin{equation}\label{identitya}
Q_L^{(a)}(z)=\sum_{n=0}^{L-1}\exp\left[z\cos\frac{2\pi}L(n+\case{1}{2})\right]
=L\sum_{l=-\infty}^\infty \cos(\pi l) I_{Ll}(z),
\end{equation}
to end up with the final expressions
\numparts
\begin{eqnarray}\label{fssa}
\fl\beta F_d^{(a)}(T,L;\phi)=&\beta F_d\left(T;\phi+\omega_0^{(a)}\right)
-
\frac2{(4\pi)^{d/2}}\left(\phi+\omega^{(a)}\right)^{\frac d2}
\sum_{l=1}^\infty (-1)^l
\frac{K_{\frac d2}\left(lL\sqrt{\phi+\omega_0^{(a)}}\right)}
{\left(\frac12lL\sqrt{\phi+\omega_0^{(a)}}\right)^{\frac d2}}\nonumber\\
\fl&
\end{eqnarray}
and
\begin{equation}
\fl K=W_d\left(\phi+\omega_0^{(a)}\right)
+\frac4{(4\pi)^{d/2}}\left(\phi+\omega^{(a)}\right)^{\frac d2-1}
\sum_{l=1}^\infty (-1)^l
\frac{K_{\frac d2-1}\left(lL\sqrt{\phi+\omega_0^{(a)}}\right)}
{\left(\frac12lL\sqrt{\phi+\omega_0^{(a)}}\right)^{\frac d2-1}}
\end{equation}
\endnumparts

For $2<d<4$, making use of the asymptotic expansion (\ref{aswb}), for large
$L$ and in the vicinity of
the bulk critical point, with $\phi+\omega_0^{(a)}<1$, we have the scaling
behaviour
\begin{equation}\label{kappaa}
\varkappa=
\frac{(y^2-\pi^2)^{\frac{d-2}2}}{(4\pi)^{d/2}}
\left[\left|\Gamma\left[\frac{2-d}2\right]\right|
-4
\sum_{l=1}^\infty (-1)^l
\frac{K_{\frac d2-1}\left(l\sqrt{y^2-\pi^2}\right)}
{(\frac12l\sqrt{y^2-\pi^2})^{\frac d2-1}}\right],
\end{equation}
where we have used $\omega_0^{(a)}\approx-\frac{\pi^2}{L^2}$.
Following the analysis of \cite{allen1993} it is easy to show that $\sqrt{\phi}$
coincides with the inverse of the finite--size correlation length $\xi_L$. In this
case,
the solution of (\ref{kappaa}) may be written as
$
\xi_L=Lf_{(a)}(\varkappa)
$,
where $f_{(a)}$ is a universal scaling function.

The critical temperature of the film corresponds to $y=\pi$ i.e.
$\phi=\left(\case{\pi}L\right)^2$, which is the asymptotic
of the lowest mode $\omega_0^{(a)}$ for large $L$. Setting $y=\pi$ in
equation (\ref{kappaa}), we find that the critical point of the film
is shifted from the bulk one by a quantity proportional to $L^{-1/\nu}$
in agreement with the finite--size scaling predictions.

For arbitrary dimension $d$, equation (\ref{kappaa}) can be solved only numerically.
%Then care must be taken in treating the cut of the square root.
Here we will specialize to the three--dimensional system, which allows
analytic treatment.
For $d=3$, equation (\ref{kappaa}) transforms into
\begin{equation}
2\pi\varkappa=\ln2\cosh\frac12\sqrt{y^2-\pi^2},
\end{equation}
whose positive solution reads
\begin{equation}
y=g_{(a)}(\varkappa),
\end{equation}
with the universal scaling function
\begin{equation}
g_{(a)}(\varkappa)=
\left(\pi^2+4[\mathrm{arccosh}\left(\case{1}{2}e^{2\pi\varkappa}\right)]^2\right)^{1/2}.
\end{equation}
At the bulk critical point, i.e. $\varkappa=0$, we obtain
\begin{equation}
y_0= g_{(a)}(0)= \frac{\sqrt{5}}3\pi.
\end{equation}
This is the critical amplitude of the finite--size correlation
length $\xi_L$. This result has
been obtained also in \cite{allen1993} using a different definition for
the correlation length imposed by the choice of a different initial equation for
the spherical field.

Using the asymptotic behaviour (\ref{uss}), valid for $2<d<4$,
and the fact that $y$ is the solution of equation
(\ref{kappaa}), we may write the
singular part of the free energy density (\ref{fabc}) in a scaling form as
\begin{eqnarray}\label{singa}
\fl f_{s,d}^{(a)}(y,\varkappa)L^{d}=&\frac12
\varkappa (y^2-\pi^2)\nonumber\\
\fl &+
\frac{(y^2-\pi^2)^{\frac d2}}{(4\pi)^{d/2}}\left(\frac1d\Gamma\left[\frac{2-d}2\right]
-2
\sum_{l=1}^\infty(-1)^l\frac{K_{\frac d2}\left(l\sqrt{y^2-\pi^2}\right)}
{\left(\frac12 l\sqrt{y^2-\pi^2}\right)^{\frac d2}}\right).
\end{eqnarray}
In accordance with the finite--size scaling hypothesis (\ref{PF}).
For the important case $d=3$, its critical amplitude at the bulk
critical temperature and
consequently the corresponding Casimir amplitude for antiperiodic
boundary conditions is\footnote{This result was obtained independently by
\cite{dantchev2008}}
\begin{equation}
\Delta^{(a)}=0.274543\ldots.
\end{equation}
Notice that the Casimir amplitude here is positive in contrast to the
case of periodic boundary conditions, see e.g. (\ref{casper}), but
approximately twice higher in magnitude. This
result is compatible with that of \cite{krech1992} obtained using
renormalization group.

Before closing this section let us mention that the expressions for
periodic boundary conditions and those corresponding to antiperiodic
boundary conditions may be written in unified general forms with a
parameter characteristic of twisted boundary conditions. In the remainder
of the paper we will use these methods to analyse the finite--size
effects of the mean spherical model with a film geometry subject to more realistic
boundary conditions.

\section{Dirichlet boundary conditions}
\subsection{Equation for the spherical field}
Here we will evaluate the finite--size contributions of the sum
appearing in (\ref{phi}) in the case of Dirichlet boundary
conditions for arbitrary $d$. We start with
\begin{equation}
\mathcal{S}_{d,L}^{(D)}(\phi)=\sum^{L}_{n=1}
W_{d-1}\left[\phi+2\cos\frac{\pi}{L+1}-2\cos\frac{\pi
n}{L+1}\right],
\end{equation}
which upon extending the sum to $n=2L+1$ may be written
\begin{eqnarray}\label{wd}
\fl\mathcal{S}_{d,L}^{(D)}(\phi)=&
\frac1{2}\sum^{2L+1}_{n=0}
W_{d-1}\left[\phi+2\cos\frac{\pi}{L+1}
-2\cos\frac{\pi n}{L+1}\right]\nonumber\\
\fl&
-\frac12W_{d-1}\left[\phi+2\cos\frac{\pi}{L+1}+2\right]
-\frac12W_{d-1}\left[\phi+2\cos\frac{\pi}{L+1}-2\right].
\end{eqnarray}
The last two terms correspond to $n=L+1$ and $n=0$,
respectively. Comparing with (\ref{spbc}) we find that the sum in
the right hand side is exactly
$\mathcal{S}_{d,2L+2}^{(p)}\left(\phi+\omega_0^{(D)}\right)$
corresponding to
periodic boundary conditions with a film of thickness $2L+2$. This suggests
that the analysis of the sum appearing in (\ref{wd}) can be performed
following the method outlined in Section \ref{periodic} (see e.g.
(\ref{spres})).

The equation for the spherical field follows from (\ref{phi}),
(\ref{wd}) and (\ref{spres}). This is (to the leading order in $L^{-1}$)
\begin{eqnarray}\label{dir}
\fl K&=W_d\left(\phi-\frac{\pi^2}{L^2}\right)
+\frac1{L}\left[W_d\left(\phi-\frac{\pi^2}{L^2}\right)
-\frac12W_{d-1}\left(\phi-\frac{\pi^2}{L^2}+4\right)
-\frac12W_{d-1}\left(\phi-\frac{\pi^2}{L^2}\right)\right]\nonumber\\
\fl&\;\;\;+\frac{4}{(4\pi)^{d/2}}
\left(\phi-\frac{\pi^2}{L^2}\right)^{\frac d2-1}
\sum_{l=1}^\infty
\frac{K_{\frac
d2-1}\left(2lL\sqrt{\phi-\frac{\pi^2}{L^2}}\right)}
{\left(lL\sqrt{\phi-\frac{\pi^2}{L^2}}\right)^{\frac d2-1}},
\end{eqnarray}
where we have used the large asymptotic behaviour
$\omega_0^{(D)}\approx-\frac{\pi^2}{L^2}$. Equation (\ref{dir}) is the
general form of the equation for the spherical field for arbitrary
dimension $d$. Notice that the right hand side is composed of a bulk
term, a size dependent surface term
\begin{equation}
W_d\left(\phi-\frac{\pi^2}{L^2}\right)
-\frac12W_{d-1}\left(\phi-\frac{\pi^2}{L^2}+4\right)
-\frac12W_{d-1}\left(\phi-\frac{\pi^2}{L^2}\right)
\end{equation}
and finite--size corrections. Here, all the quantities are function of the
combination $\phi-\frac{\pi^2}{L^2}$.
Thus, the critical properties of the mean spherical model of
finite thickness under Dirichlet boundary conditions should be
investigated using $\phi-\frac{\pi^2}{L^2}$ as a small parameter
keeping in mind that $\phi\ll1$ and $L\gg1$.

The scaling behaviour of (\ref{dir}) depends strongly upon $d$,
indeed the asymptotic expansion of $W_{d-1}(z)$ for small argument
takes different expressions for different values of $d$. We first
start with the important three dimensional case that has been
extensively studied in the literature. Later we will extend our
analysis to the interval $3<d<4$. This constraint for $d$ ensures the
validity of the asymptotic expansion (\ref{aswb}) for $W_d(z)$ and
$W_{d-1}(z)$ at the same time.
%This means that the lower critical dimension in this case is $d=3$.

For $d=3$, using (\ref{aswb}) with $d=3$ and
\begin{equation}
W_2(z)= \frac1{4\pi}\left(5\ln2-\ln z\right)+O(z\ln z),
\end{equation}
form (\ref{dir}) we have
\begin{equation}\label{dir3}
\fl K-K_{c,3}=\frac1L\left[K_{c,3}-\frac12W_2(4)-\frac{5\ln2}{8\pi}\right]
-\frac1{4\pi L}\ln L
-\frac1{4\pi
L}\ln\frac{\sinh2\sqrt{L^2\phi-\pi^2}}{\sqrt{L^2\phi-\pi^2}},
\end{equation}
which, apart from an unimportant numerical factor that enters in the definition
of $K$, coincides with equation (4.69) of \cite{barber1973}
obtained at $d=3$ via a different method.
Consequently our result disagrees with that of \cite{chen2003},
where it has been argued that equation (4.69) of \cite{barber1973} was
incorrect far from $K_{c,3}$\footnote{The origin of this discrepancy is
discussed in \cite{chen2003}.}.
At the bulk critical point, solving (\ref{dir3}), with the assumption
$L\sqrt{\phi}\ll1$, we find $\phi\sim L^{-3}$ in
agreement with \cite{chen2003,dantchev2003}. It is worth
mentioning that (\ref{dir3}) cannot be put in a scaling form
because of the logarithmic dependence on $L$. However, if one
considers a shifted critical point that absorbs the term proportional
to $\ln L$ and the surface contributions according to \cite{barber1973}
\begin{equation}\label{stcd3}
K_{s,3}^{(D)}=K_{c,3}+
\frac1L\left[K_{c,3}-\frac12W_2(4)-\frac{7\ln2}{8\pi}\right]
-\frac1{4\pi L}\ln L,
\end{equation}
it is possible to recover the scaling behaviour.
%Unfortunately, the shifted critical temperature (\ref{stcd3}) has no physical meaning,
Notice that expression (\ref{stcd3}) does not conform with the
predictions of the theory of finite--size scaling for the finite--size
shift of the critical temperature temperature,
%although at first sight one would
%hope to write $L^{-1}$ as $L^{-1/\nu}$ with ($\nu=(d-2)^{-1}$), which is
%actually not correct as we will see later for $3<d<4$.
%Furthermore,
since there is a term containing $\ln L$.
%also violates these predictions.

In the case $3<d<4$, using the asymptotic expansion
(\ref{aswb}), the equation for the spherical field (\ref{dir})
may be written as
\begin{eqnarray}\label{scadd}
\fl-\varkappa=&L^{d-3}\left[W_d(0)-\case12W_{d-1}(0)-\case12W_{d-1}(4)\right]
\nonumber\\
\fl&+\frac1{(4\pi)^{d/2}}\left(y^2-\pi^2\right)^{\frac d2-1}
\left[\Gamma\left(\frac{2-d}2\right)
-\sqrt{\pi}\Gamma\left(\frac{3-d}2\right)\left(y^2-\pi^2\right)^{-\frac12}\right]\nonumber\\
\fl&+\frac1{(4\pi)^{d/2}}\left(y^2-\pi^2\right)^{\frac d2-1}
\sum_{l=1}^\infty
\frac{K_{\frac
d2-1}\left(2l\sqrt{y^2-\pi^2}\right)}
{\left(l\sqrt{y^2-\pi^2}\right)^{\frac d2-1}},
\end{eqnarray}
where we have introduced the usual notations
$\varkappa=L^{1/\nu}(K_{c,d}-K)$ and
$y=L\sqrt{\phi}$.
The first term in the right hand side shows that the standard finite--size scaling
hypothesis breaks down in the vicinity of the bulk critical
temperature. It is likely that the introduction of a
scaling function depending on $\varkappa$ and an additional variable
to take care of the term proportional to
$L^{d-3}$ aiming at the modification of the finite--size scaling
would cure this deficiency.

From equation (\ref{dir}) it is easy to see that leading
large $L$ behaviour of the finite--size shift from the bulk critical
temperature to the one where the film is expected to have a singular
behaviour is $L^{-1}$. This result, valid for any dimension, shows that
the finite--size scaling hypothesis is violated as it has been pointed
out by \citeasnoun{brezin1983}.

\subsection{Free energy density}
Let us now turn to the evaluation of the free energy density
(\ref{freee}). We need
the asymptotic behaviour of the sum
\begin{equation}
\mathcal{P}^{(D)}_{d,L}(\phi)=\sum_{n=1}^LU_{d-1}
\left[\phi+2\cos\frac{\pi}{L+1}-2\cos\frac{\pi n}{L+1}\right],
\end{equation}
entering the expression (\ref{freee}) of the free energy.
Here again we extend the sum to $n=2L+1$, to get
\begin{eqnarray}\label{ud}
\fl\mathcal{P}^{(D)}_{d,L}(\phi)=&\frac1{2}\sum_{n=0}^{2L+1}U_{d-1}
\left[\phi+2\cos\frac{\pi}{L+1}-2\cos\frac{\pi n}{L+1}\right]
\nonumber\\
\fl&-\frac1{2}U_{d-1}
\left[\phi+2\cos\frac{\pi}{L+1}+2\right]
-\frac1{2}U_{d-1}
\left[\phi+2\cos\frac{\pi}{L+1}-2\right].
\end{eqnarray}

The sum in the right hand side of (\ref{ud}) corresponds to
$P_{d,2L+2}^{(p)}\left(\phi+\omega_0^{(D)}\right)$ from
(\ref{pper}). Then from (\ref{pp}) and (\ref{ud}) we get
the free energy density
\begin{eqnarray}\label{fend}
\beta F_d^{(D)}(T,L;\phi)&=&\beta F_d\left(T;\phi-\frac{\pi^2}{L^2}\right)
+\frac1L\beta F_{d,{\mathrm{surf.}}}^{(D)}\left(T;\phi-\frac{\pi^2}{L^2}\right)\nonumber\\
&&
-\frac2{(4\pi)^{\frac d2}}\left(\phi-\frac{\pi^2}{L^2}\right)^{\frac d2}
\sum_{l=1}^\infty\frac{ K_{\frac d2}
\left(2lL\sqrt{\phi-\frac{\pi^2}{L^2}}\right)}
{\left(lL\sqrt{\phi-\frac{\pi^2}{L^2}}\right)^{\frac d2}},
\end{eqnarray}
where
$$
\beta F_{d,\mathrm{surf.}}^{(D)}(T;\sigma)=\frac12
\left[U_d(\sigma)-
\frac12U_{d-1}(\sigma+4)
-\frac12U_{d-1}(\sigma)\right]
$$
accounts for size dependent contributions stemming from the
surfaces. In the following we will discuss the finite--size
behaviour of the free energy and its dependence on the dimensionality
$d$. Again we discuss separately the cases $d=3$ and $3<d<4$ imposed by the
validity of the expansion (\ref{uss}) for $U_d(z)$ and $U_{d-1}(z)$,
simultaneously.

For $d=3$, we use the expansions (\ref{uss}) and
\begin{equation}\label{usp}
U_2(z)=U_2(0)-\frac1{4\pi}z\ln z
+\frac1{4\pi}(1+5\ln2)z +O(z^2\ln z)
\end{equation}
to obtain the singular part of the free energy density
\begin{eqnarray}\label{sind}
\fl\beta f_{s,3}^{(D)}L^3=&
\frac12\varkappa\left(y^2-\pi^2\right)
-\frac1{12\pi}\left(y^2-\pi^2\right)^{\frac32}
+\frac1{2}\left[K_{c,3}-\case{1}{2} W_2(4)-\frac{1+5\ln2}{8\pi}\right]
\left(y^2-\pi^2\right)
\nonumber\\
\fl&
+\frac1{16\pi}\left(y^2-\pi^2\right)\ln\left(y^2-\pi^2\right)
-\frac1{8\pi}\left(y^2-\pi^2\right)\ln L
\nonumber\\
\fl&
-\frac1{8\pi}\sqrt{y^2-\pi^2}
\ \mathrm{Li}_2\left(\exp\left[-2\sqrt{y^2-\pi^2}\right]\right)
-\frac1{16\pi}
\mathrm{Li}_3\left(\exp\left[-2\sqrt{y^2-\pi^2}\right]\right)\nonumber\\
\fl&
\end{eqnarray}
At the shifted critical point $K_{s,3}^{(D)}$ (See (\ref{stcd3})),
$f_{s,3}^{(D)}$ takes a scaling form, as it has been pointed out
by \citeasnoun{barber1974a}.
Notice that at the bulk critical temperature
$f_{s,3}^{(D)}$ is proportional to $L^{-3}\ln L$ implying that the finite
size--scaling hypothesis (\ref{PF}) breaks down
for the mean spherical model with Dirichlet boundary conditions at
$d=3$, while in more realistic models the finite--size scaling is
valid and the critical Casimir amplitude can be estimated \cite{krech1992}.

Another expression for the scaling behaviour of the singular
part of free energy density was obtained in \cite{barber1974a}.
In our notations it reads
\begin{equation}\label{barber}
\fl\beta f_{s,3}^{B}L^3=\frac12\varkappa y^2-\frac{y^2}{8\pi}\ln L
+\frac{y^2}2\left[K_{c,3}-\case{1}{2}W_2(4)-\frac{7\ln2}{8\pi}\right]
+Q_0(y),
\end{equation}
where
\begin{equation}\label{intrepb}
Q_0(x)=\frac\pi8\left[R(-1)-R\left(\frac{x^2}{\pi^2}-1\right)\right]
\end{equation}
with
$$
R(z)=\int_0^z\ln\frac{\sinh\pi\sqrt{w}}{\pi\sqrt{w}}dw.
$$

\begin{figure}[th!]
\begin{center}
\resizebox{.8\columnwidth}{!}{\includegraphics{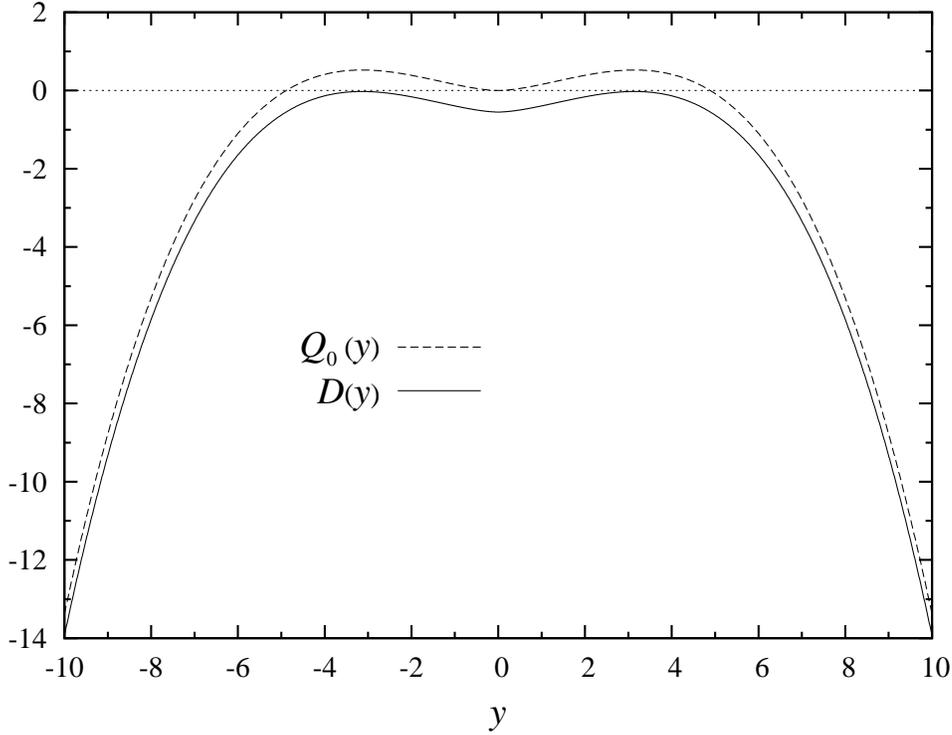}}
\end{center}
\caption{Behaviours of the functions $Q_0(y)$ (\ref{barber}) and
$D(y)$ (\ref{fd}) against $y$.}
\label{scale}
\end{figure}

By comparing (\ref{sind}) and (\ref{barber}) we see that
the first three terms in the right hand side of (\ref{barber}) have
their counterparts in (\ref{sind}) obtained through the replacement
$y^2\to y^2-\pi^2$. It seems that the terms linear in $\pi^2$ and the
term proportional to $L^{-3}\ln L$ were
neglected in (\ref{barber}).
It remains to see what is the situation for the rest of the terms.
Unfortunately no direct analytic comparison can be made, despite the
fact that the integral representation (\ref{intrepb}) can be expressed in
terms of polylogarithms. To achieve the comparison recourse must be sought
in numerical methods, so we plot the function $Q_0(y)$ and
\begin{eqnarray}\label{fd}
\fl D(y)=&
\frac1{16\pi}(2\ln2-1)\left(y^2-\pi^2\right)
-\frac1{12\pi}\left(y^2-\pi^2\right)^{\frac32}
+\frac1{16\pi}\left(y^2-\pi^2\right)\ln\left(y^2-\pi^2\right)
\nonumber\\
\fl&
-\frac1{8\pi}\sqrt{y^2-\pi^2}
\ \mathrm{Li}_2\left(\exp\left[-2\sqrt{y^2-\pi^2}\right]\right)
-\frac1{16\pi}
\mathrm{Li}_3\left(\exp\left[-2\sqrt{y^2-\pi^2}\right]\right)\nonumber\\
\fl&
\end{eqnarray}
from (\ref{sind}) which contains terms that are not present in
(\ref{barber}) and those that were apparently neglected.
The result is shown in Figure \ref{scale}.\footnote{The
numerical evaluation of the named expressions was performed
with the aid of WOLFRAM MATHEMATICA 6}
We see clearly
that both functions have similar behaviours and are shifted one from
the other by a constant. This has been checked by computing the
derivatives of both functions which gives us the same result. The
difference between the two functions is estimated to be
\begin{equation}
Q_0(y)-D(y)=\frac\pi{16}\left[-1+\ln(4\pi^2)+\frac1{\pi^2}\zeta(3)\right].
\end{equation}
Thus, the scaling functions (\ref{sind}) and (\ref{barber}) are equal
up to an irrelevant constant that does not become singular at the bulk
critical point.

For $3<d<4$, using the asymptotic expansion (\ref{uss}),
the scaling form of the singular part of the free energy
density (\ref{fend}) reads
\begin{eqnarray}
\fl 2f_{s,d}^{(D)}L^d=&\varkappa\left(y^2-\pi^2\right)
+L^{d-3}\left[W_d(0)-\case12W_{d-1}(0)-\case12W_{d-1}(4)\right]\left(y^2-\pi^2\right)
\nonumber\\
\fl&
+\frac2{(4\pi)^{\frac d2}}\left(y^2-\pi^2\right)^{\frac d2}
\left[\frac1d\Gamma\left(\frac{2-d}2\right)-\frac{\sqrt{\pi}}{d-1}
\Gamma\left(\frac{3-d}2\right)\frac1{\sqrt{y^2-\pi^2}}\right]
\nonumber\\
\fl&
-\frac2{(4\pi)^{\frac d2}}\left(y^2-\pi^2\right)^{\frac d2}
\sum_{l=1}^\infty\frac{ K_{\frac d2}
\left(2l\sqrt{y-\pi^2}\right)}
{\left(l\sqrt{y^2-\pi^2}\right)^{\frac d2}}.
\end{eqnarray}
Just like the equation for the spherical field we find that the surface
contributions to the singular part of the free energy violates the
standard finite--size scaling hypothesis (\ref{PF}).
A reformulation of the scaling behaviour would require the
introduction of a function with two arguments: the scaling variable
$\varkappa$ and a
scaling variable that incorporates the size dependent term $L^{d-3}$.

\section{Neumann boundary conditions}\label{sNeumann}
\subsection{Equation for the spherical field}
To investigate the finite--size effects in the mean spherical model of
finite thickness subject to Neumann boundary conditions,
through equation (\ref{phi}),
we need to estimate the large asymptotic behaviour of the sum
\begin{eqnarray}\label{wn}
\fl\mathcal{S}_{d,L}^{(N)}(\phi)&=\sum^{L}_{n=1}
W_{d-1}\left[\phi+2-2\cos\frac{\pi(n-1)}L\right]\nonumber\\
\fl&=\frac1{2}\sum_{n=0}^{2L-1}W_{d-1}
\left[\phi+2-2\cos\frac{\pi n}L\right]
-\frac1{2}W_{d-1}\left[\phi+4\right]
+\frac1{2}W_{d-1}\left[\phi\right].
\end{eqnarray}
The sum in the last line is exactly $\mathcal{S}_{d,2L}^{(N)}(\phi)$ (see
(\ref{spbc})).
Consequently, from (\ref{wn}) and (\ref{spres}), the equation for
the spherical field is given by
\begin{equation}\label{phin}
\fl K=W_d(\phi)
-\frac1{2L}\left[
W_{d-1}(\phi+4)
-W_{d-1}(\phi)\right]+
\frac{4}{(4\pi)^{d/2}}\phi^{\frac d2-1}
\sum_{l=1}^\infty\frac{K_{\frac d2-1}\left(2lL\sqrt{\phi}\right)}
{\left(lL\sqrt{\phi}\right)^{\frac d2-1}},
\end{equation}
to the leading order in $L^{-1}$.

In addition to the restriction on $d$ imposed by the fact that $d=2$
is the lower critical dimension and $d=4$ is the upper one, which is
contained in $W_d(z)$ there is another one originating from
$W_{d-1}(z)$ that restricts the validity of the asymptotic
behaviour (\ref{aswb}) to the interval $3<d<4$. For that reason we will
investigate in some details only the cases $d=3$ and $3<d<4$. The
remaining part of the interval i.e. $2<d<3$ requires special
treatment.

%The presence of $W_{d-1}(\phi)$ in (\ref{phin}) raises the lower
%critical dimension of the film to $d=3$.
At $d=3$,
equation (\ref{phin}) takes the simple form
\begin{equation}\label{scnd3}
\fl K-K_{c,3}=-\frac1{2L}\left[W_2(4)-\frac5{4\pi}\ln2\right]
+\frac1{4\pi L}\ln L-\frac1{4\pi L}\ln\left[2L\sqrt{\phi}
\sinh(L\sqrt{\phi})\right].
\end{equation}
This equation was derived in \cite{dantchev2003} using a
method based on
\cite{barber1973}. At the bulk critical temperature $K_{c,3}$ the
spherical field behaves as $\phi\sim L^{-1}$ to the leading order
assuming $L\sqrt{\phi}\ll1$. In the limit $L\sqrt{\phi}\gg1$ one would
expect a logarithmic behaviour as pointed out by \cite{dantchev2003}.
At the shifted critical temperature, defined through
\cite{dantchev2003}
\begin{equation}\label{stcn3}
K_{s,3}^{(N)}=K_{c,3}-\frac1{2L}\left[W_2(4)-\frac3{4\pi}\ln2\right]
+\frac1{4\pi L}\ln L,
\end{equation}
equation (\ref{scnd3}) may be written in a scaling form.
An inspection of the expression (\ref{stcn3}) shows that
the predictions of the theory of finite--size scaling for the shifted
critical temperature is not fulfilled due to the
presence of the term proportional to $\ln L$.
%due to the fact that $d=3$ is the lower critical dimension.

For $3<d<4$ the scaling form of the equation for the spherical field
(\ref{phin}) is given by
\begin{eqnarray}\label{scnd}
\fl\varkappa =& \frac12L^{d-3}\left[W_{d-1}(4)-W_{d-1}(0)\right]
-\frac1{(4\pi)^{d/2}}\left[\Gamma\left(\frac{2-d}2\right)
-\frac{\sqrt{\pi}}y\Gamma\left(\frac{3-d}2\right)\right]y^{d-2}\nonumber\\
\fl &
-
\frac{4}{(4\pi)^{d/2}}y^{d-2}
\sum_{l=1}^\infty\frac{K_{\frac d2-1}\left(2ly\right)}
{\left(ly\right)^{\frac d2-1}}.
\end{eqnarray}
Here again we used the notations $\varkappa=L^{-1/\nu}(K_{c,d}-K)$
and $y=L\sqrt{\phi}$.
As above for the case of Dirichlet boundary conditions the standard finite--size
scaling hypothesis is violated here as well and would need a
reformulation in order to take into account the term proportional to
$L^{d-3}$. Furthermore, equation (\ref{scnd}) shows that the leading
asymptotic behaviour of the finite--size shift of the critical temperature is
$L^{-1}$, which is not consistent with the predictions of the theory
of finite--size scaling.

\subsection{Free energy density}
The free energy density (\ref{freee}) in the case of Neumann
boundary conditions is obtained by analysing the sum
\begin{eqnarray}\label{sn}
\fl \mathcal{P}_{d,L}^{(N)}(\phi)&=\sum^{L}_{n=1}
U_{d-1}\left[\phi+2-2\cos\frac{\pi(n-1)}L\right]\nonumber\\
\fl &=\frac1{2}\sum_{n=0}^{2L-1}U_{d-1}
\left[\phi+2-2\cos\frac{\pi n}L\right]
-\frac1{2}U_{d-1}\left[\phi+4\right]
+\frac1{2}U_{d-1}\left[\phi\right].
\end{eqnarray}
which contains the term $\mathcal{P}_{d,2L}^{(p)}(\phi)$, equivalent to (\ref{pper}).
Using (\ref{sn}) and (\ref{pp}) we obtain explicitly
\begin{equation}
\fl\beta F_d^{(N)}(T,L;\phi)=\beta F_d\left(T;\phi\right)
+\frac1L\beta F_{d,\mathrm{surf.}}^{(N)}(\phi)
-\frac2{(4\pi)^{\frac d2}}\phi^{\frac d2}
\sum_{l=1}^\infty\frac{ K_{\frac d2}
\left(2lL\sqrt{\phi}\right)}
{\left(lL\sqrt{\phi}\right)^{\frac d2}},
\end{equation}
where
\begin{equation}
\beta F_{d,\mathrm{surf.}}^{(N)}(\phi)=
\frac1{4}U_{d-1}\left[\phi\right]
-\frac1{4}U_{d-1}\left[\phi+4\right]
\end{equation}
is the surfaces contribution.

Again we are faced with a situation where we have to restrict the
validity of our expressions due the expansion (\ref{uss}), which
should hold for $U_d(z)$ and $U_{d-1}(z)$ simultaneously. Here we
consider the cases $d=3$ and $3<d<4$ separately.

At $d=3$, using (\ref{uss}) and (\ref{usp})
the singular part of the free energy density (\ref{sn}) reads
\begin{eqnarray}
\fl\beta f_{d,3}^{(N)}L^3=&\frac12\varkappa y^2-\frac1{12\pi}y^{3}
-\frac14\left[W_2(4)-\frac{1+5\ln2}{4\pi}\right]y^2
-\frac1{8\pi}y^2\ln y+\frac1{8\pi}y^2\ln L
\nonumber\\
\fl&
-\frac1{8\pi}y\mathrm{Li}_2(e^{-2y})
-\frac1{16\pi}y\mathrm{Li}_3(e^{-2y}).
\end{eqnarray}
Here again we have a logarithmic dependence on $L$ and surface
contributions leading to the violation of the finite--size
scaling hypothesis (\ref{PF}).
%At the shifted critical point (\ref{stcn3}) the scaling is recovered.
This makes the
spherical model unsuitable for the evaluation of the Casimir amplitude
for $O(n)$ systems with Neumann boundary conditions. The value of this quantity is
known from renormalization group \cite{grueneberg2008}.

For $3<d<4$, the singular part of the free energy reads
\begin{eqnarray}
\fl \beta f_{s,d}^{(N)}L^d=&\frac12\varkappa y^2+\frac14L^{d-3}
\left[W_{d-1}(0)-W_{d-1}(4)\right]y^2\nonumber\\
\fl &+
\frac1{2(4\pi)^{d/2}}\left[\frac1d\Gamma\left(\frac{2-d}2\right)
+\frac{\sqrt{\pi}}{d-1}\Gamma\left(\frac{3-d}2\right)\right]
-\frac2{(4\pi)^{d/2}}y^d
\sum_{l=1}^\infty\frac{K_\frac d2(2ly)}{(ly)^\frac d2}.
\nonumber\\\fl&
\end{eqnarray}
This suggests that standard finite--size scaling hypothesis (\ref{PF})
is not valid for
the mean spherical model confined to a film geometry with Neumann
boundary conditions.
%{\blue If the term proportional to $L^{d-3}$ is
%absorbed in the definition of the distance to the critical point
%of the film (\ref{stcnd}), then the scaling behaviour is obtained.}
Similar to the case of Dirichlet boundary conditions a modified finite--size
scaling hypothesis is necessary to get the appropriate scaling
behaviour.

\section{Neumann--Dirichlet boundary conditions}\label{smixed}
\subsection{Equation for the spherical field}
To extract the finite--size effects from (\ref{phi}) in the case of
mixed boundary conditions i.e. Neumann and Dirichlet on the bounding surfaces,
the sum to be considered is
\begin{eqnarray}\label{sm}
\fl\mathcal{S}_{d,L}^{(DN)}(\phi)&=\sum_{n=1}^{L}
W_{d-1}\left[\phi+2\cos\frac{\pi}{2L+1}
-2\cos\frac{2\pi\left(n-\frac12\right)}{2L+1}\right]\nonumber\\
\fl&=\frac1{2}\sum_{n=0}^{2L}
W_{d-1}\left[\phi+2\cos\frac{\pi}{2L+1}
-2\cos\frac{2\pi\left(n+\frac12\right)}{2L+1}\right]
\nonumber\\
\fl&\;\;\;\;
-\frac1{2}W_{d-1}\left[\phi+2\cos\frac{\pi}{2L+1}
+2\right].
\end{eqnarray}
By inspection of the sum in the second line we see that the analysis
of this case is tightly related to that of antiperiodic boundary
conditions. In other words this sum is exactly
$S^{(a)}_{d,2L+1}(\phi+\omega_0^{(ND)})$ with $S^{(a)}_{d,2L+1}(\sigma)$
defined in (\ref{sapbc}) meaning that in this case the finite--size
corrections correspond to a film of
thickness $2L+1$ subject to antiperiodic boundary conditions.
Then, along lines similar to the analysis of Section \ref{antiperiodic}, for the
equation for the spherical field we get, for arbitrary $d$,
\begin{eqnarray}
\fl K=&W_d\left(\phi-\frac{\pi^2}{4L^2}\right)
+\frac1{2L}\left[W_d\left(\phi-\frac{\pi^2}{4L^2}\right)
-W_{d-1}\left(\phi-\frac{\pi^2}{4L^2}+4\right)\right]
\nonumber \\
\fl&+\frac4{(4\pi)^{d/2}}\left(\phi-\frac{\pi^2}{4L^2}\right)^{\frac
d2-1}\sum_{k=1}^\infty(-1)^l
\frac{K_{\frac d2-1}\left(2lL\sqrt{\phi-\frac{\pi^2}{4L^2}}\right)}
{\left(lL\sqrt{\phi-\frac{\pi^2}{4L^2}}\right)^{\frac d2-1}}
\end{eqnarray}
to the leading order in $L^{-1}$.
%Notice that here the lower critical
%dimension of the film is not altered i.e. it coincides with that of the
%bulk system.
As before we will restrict ourselves between the lower and the upper critical
dimensions.
For $2<d<4$, using (\ref{aswb}), we have
\begin{eqnarray}\label{kappam}
\fl\varkappa=&
\frac{\left(y^2-\frac{\pi^2}4\right)^{\frac{d-2}2}}{(4\pi)^{d/2}}
\left[\left|\Gamma\left[\frac{2-d}2\right]\right|
-4
\sum_{l=1}^\infty (-1)^l
\frac{K_{\frac d2-1}\left(2l\sqrt{y^2-\frac{\pi^2}4}\right)}
{\left(l\sqrt{y^2-\frac{\pi^2}4}\right)^{\frac d2-1}}\right]\nonumber\\
\fl&-\frac12L^{d-3}\left[K_{c,d}
-W_{d-1}(4)\right].
\end{eqnarray}
With $y=L\sqrt{\phi}$ and $\varkappa=L^{1/\nu}(K_{c,d}-K)$.
The solution of this equation depends \textit{explicitly} on the size
$L$ which leads us to the
conclusion that the standard finite--size scaling is violated here as
well. To resolve this issue one would need a modified finite--size
scaling assumption. This remains valid also for the
corresponding finite--size shift of the critical temperature whose
leading asymptotic behaviour is $L^{-1}$, rather the than the predicted
$L^{-1/\nu}$.
%{\blue
%Let us introduce the critical temperature of the film via
%\begin{equation}\label{kcndd}
%K_{c,d}^{(ND)}=K_{c,d}-\frac1{2L}\left[K_{c,d}-W_{d-1}(4)\right].
%\end{equation}
%In this case it is possible to write (\ref{kappam}) in a scaling form.
%}
Remark however that for $d=3$ the term $L^{d-3}$ vanishes and the
finite--size scaling is expected to hold. Indeed,
at $d=3$, (\ref{kappam}) turns into the simple form
\begin{equation}\label{mixed3d}
4\pi\varkappa=\ln2\cosh\sqrt{y^2-\frac{\pi^2}4}
-2\pi\left[K_{c,3}-W_{2}(4)\right],
\end{equation}
This equation is identical to equation (4.31) of \cite{dantchev2003} obtained
by a method adapted after \cite{barber1973}.
The positive solution of (\ref{kappam}) may be written in the scaling form
\begin{equation}
y=g_{(ND)}(\varkappa),
\end{equation}
where the universal scaling function is given by
\begin{equation}
\fl g_{(ND)}(\varkappa)=
\left[\frac{\pi^2}4+\left(\mathrm{arccosh}
\left[\case{1}{2}\exp\left(4\pi\varkappa+
2\pi\left[K_{c,3}-W_{2}(4)\right]\right)
\right]\right)^2\right]^{1/2}.
\end{equation}
At the bulk critical temperature, $\varkappa=0$,
%solving (\ref{mixed3d}) numerically
we find the critical amplitude
\begin{equation}
y_0=g_{(ND)}(0)=1.45684\ldots.
\end{equation}

\subsection{Free energy density}
The free energy density from equation (\ref{freee}) in the
case of Neumann--Dirichlet
is obtained as a result of the analysis of the sum
\begin{eqnarray}\label{pm}
\fl\mathcal{P}_{d,L}^{(DN)}(\phi)&=\sum_{n=1}^{L}
U_{d-1}\left[\phi+2\cos\frac{\pi}{2L+1}
-2\cos\frac{2\pi\left(n-\frac12\right)}{2L+1}\right]\nonumber\\
\fl&=\frac1{2}\sum_{n=0}^{2L}
U_{d-1}\left[\phi+2\cos\frac{\pi}{2L+1}
-2\cos\frac{2\pi\left(n+\frac12\right)}{2L+1}\right]
\nonumber\\
\fl&\;\;\;\;
-\frac1{2}U_{d-1}\left[\phi+2\cos\frac{\pi}{2L+1}
+2\right].
\end{eqnarray}
So apart from a surface term (the last term) the asymptotic form of
the free energy has
a similar expression to (\ref{fssa})
in the case of the antiperiodic boundary conditions with the
replacements $\omega_0^{(a)}\to\omega_0^{(ND)}$ and $L\to2L$ i.e.
\begin{eqnarray}
\fl\beta F_d^{(ND)}(T,L;\phi)=&\beta F_d\left(T;\phi+\omega_0^{(ND)}\right)
+\frac1L\beta F_{d,\mathrm{surf.}}^{(ND)}(\phi+\omega_0^{(ND)})
\nonumber\\
\fl&
-
\frac2{(4\pi)^{d/2}}\left(\phi+\omega^{(ND)}\right)^{\frac d2}
\sum_{l=1}^\infty (-1)^l
\frac{K_{\frac d2}\left(2lL\sqrt{\phi+\omega_0^{(ND)}}\right)}
{\left(lL\sqrt{\phi+\omega_0^{(ND)}}\right)^{\frac d2}},
\end{eqnarray}
with the surface contribution
\begin{equation}
F_{d,\mathrm{surf.}}^{(ND)}(\sigma)=
\frac14\left[U_d(\sigma)-U_{d-1}(\sigma+4)\right].
\end{equation}

For $2<d<4$, the singular part of the free energy density follows also from that
for antiperiodic boundary conditions (\ref{singa}) and the corrections
originating from the surfaces. In this case we get
\begin{eqnarray}\label{singm}
\fl\beta f_{s,d}^{(ND)}(y,\varkappa)=&\frac12L^{-d}
\varkappa\left(y^2-\frac{\pi^2}4\right)
+\frac1{4}L^{-3}\left[K_{c,d}
-W_{d-1}(4)\right]\left(y^2-\frac{\pi^2}{4}\right)
\nonumber\\
&+
L^{-d}\frac{(y^2-\frac{\pi^2}4)^{\frac d2}}{(4\pi)^{d/2}}
\left(\frac1d\Gamma\left[\frac{2-d}2\right]
-2
\sum_{l=1}^\infty(-1)^l\frac{K_{\frac
d2}\left(2l\sqrt{y^2-\frac{\pi^2}4}\right)}
{\left(l\sqrt{y^2-\frac{\pi^2}4}\right)^{\frac d2}}\right).
\nonumber\\
\fl&
\end{eqnarray}
Here again there is a term proportional to $L^{d-3}$ that violates the
standard finite--size scaling hypothesis (\ref{PF}) in the vicinity of $K_{c,d}$
and its modification is necessary to get the correct scaling.
%{\blue
%At the film critical point $K_{c,d}^{(ND)}$ from (\ref{kcndd})
%we may write the free energy in a scaling form.
%}
At $d=3$ there is no explicit dependence in $L$.
In this case, the critical amplitude of the
singular part of the free energy density is obtained from
(\ref{singm}). Thus the Casimir amplitude for the
three dimensional mean spherical film subject to Neumann--Dirichlet
boundary conditions is
\begin{equation}
\Delta^{(ND)}=0.01922\ldots.
\end{equation}
This exact result is in conformity with that of \cite{krech1992}
obtained using renormalization group.
Similar to the case of antiperiodic boundary conditions the Casimir
amplitude is positive but smaller in magnitude indicating a weaker
repulsing force between the surfaces bounding the system.

\section{Discussion}\label{discussion}

We investigated the finite--size effects in the $d$ dimensional
ferromagnetic mean spherical model of finite thickness $L$ subject to
different kinds of boundary conditions: periodic $(p)$, antiperiodic
$(a)$,
Dirichlet $(D)$ and Neumann $(N)$ on both surfaces bounding the model, and a
combination of Neumann and Dirichlet on each surface $(ND)$.
We proposed a method for the computation of the finite--size
corrections of the free energy for arbitrary dimension. Our analysis
showed that for Dirichlet and Neumann boundary conditions the
finite--size effects are essentially equivalent to the case of periodic
boundary conditions for a film of thickness $2L$ and
additional surface terms. Similarly, the case of Neumann--Dirichlet
was found to be related to the case of antiperiodic boundary
conditions with thickness $2L$.

The free energy density and the equation for the spherical field were
computed for a film with arbitrary dimension $d$ subject to the different
boundary conditions. In the particular case $d=3$, our general
expressions for $(D)$, $(N)$ and $(ND)$ reduce to those
obtained by \cite{barber1973} and
\cite{danchev1997,dantchev2003}. It is found that the singular part of
the free energy density has the standard finite--size scaling form for
$2<d<4$ only in the cases $(p)$ and $(a)$ i.e.
for those boundary conditions which do not break the translation
invariance of the model. In these cases we estimated the critical
amplitude of the singular part of the free energy
and obtained the
values: $\Delta^{(p)}=-2\zeta(3)/(5\pi)=-0.153051\ldots$
and $\Delta^{(a)}=0.274543\ldots$ for $(p)$ and $(a)$, respectively.
Interpreted in terms of the Casimir effect this imply
that in the case $(p)$ we have fluctuation--induced attraction between
the surfaces bounding the model and a repulsion in the case $(a)$.

For a film subject to (D) or (N)
%the lower critical dimension is raised to 3
the critical point of the film is shifted from the bulk one by
surface and finite--size terms, whose leading asymptotic behaviour
is proportional to $L^{-1}$. This results disagrees
with the scaling behaviour of the finite--size shift predicted by the
theory of finite--size scaling. In
the vicinity of the bulk critical point the standard finite--size scaling
is not valid in general.
At $d=3$, the solution of the equation for the spherical field is
proportional to $L^{-3}$ for $(D)$ and to $L^{-1}$ for $(N)$, while
the singular part of the free energy exhibits a logarithmic dependence
on $L$.
For $3<d<4$, the standard finite--size scaling for the
singular part of the free energy needs to be modified.
%according to
%\begin{equation}\label{mod}
%f^{(\tau)}_{s,d}(t;L)\approx
%L^{-d}Y^{(\tau)}(b_1tL^{1/\nu},b_2L^{d-3}),
%\end{equation}
%with $b_1$ and $b_2$ some non--universal quantities depending on the
%details of the model. The function $Y^{(\tau)}$ is expected to be a
%universal function of its arguments.

In the case $(ND)$,
%the lower critical dimension is the same as that of the bulk model. Nonetheless,
the film has its own critical point
shifted from the bulk one by surface terms, as well as finite--size
effects. The finite--size shift of the critical temperature and the
scaling form of the free energy do not conform with the theory of
finite--size scaling. For $2<d<4$, the standard
finite--size scaling has to be modified in the neighbourhood of the bulk
critical temperature. %using (\ref{mod}).
A surprising fact is that the finite--size
scaling hypothesis is again valid at $d=3$. At $T_c$,
the square root of the equation for the spherical field is equal to
$1.45684\ldots L^{-1}$ and
the Casimir amplitude is found to
be $\Delta^{(ND)}=0.01922\ldots$ i.e. a weaker than the case $(a)$
fluctuation--induced repulsion.

It is well known that the mean spherical model is not able to capture the
gross features of $O(n)$ models when it is subject to
boundary conditions that break the translational invariance.
%of the model.
To solve this problem remaining in the framework of the
spherical model one would introduce additional spherical fields to
ensure the proper behaviour of the surface spins \cite{singh1975}.
Otherwise one can try recovering the equivalence between the spherical
model and $O(n)$ spin models by imposing spherical constraints
ensuring the same mean square value for all spins of the
system \cite{knops1973}. In the case of a film geometry this is
equivalent to having a spherical constraint on each layer of the
system with a space dependent spherical field along the finite
direction
\cite{hikami1976,bray1977,ohno1983} whose relaxed version reduces to
the model under consideration. Even in this case an
accurate study of the problems related to finite--size scaling remains
rather untractable analytically (see e.g. \cite{hikami1976,brezin1983}.

\ack
The author expresses his gratitude to Prof. Diehl for many
illuminating discussions and in addition to Prof. Tonchev for a
critical reading of the manuscript.

%\bibliographystyle{ieeetr}
%\bibliography{../fssdyn/biblio}
%\bibliography{biblio}

\end{document}